\documentclass[usenatbib]{mnras}
\usepackage{graphicx}
\usepackage{amsmath}
\usepackage{amssymb}
\usepackage{color}
\usepackage{url}

\usepackage{threeparttable}

\newcommand\lsim{\mathrel{\rlap{\lower4pt\hbox{\hskip1pt$\sim$}}
        \raise1pt\hbox{$<$}}}
\newcommand\gsim{\mathrel{\rlap{\lower4pt\hbox{\hskip1pt$\sim$}}
        \raise1pt\hbox{$>$}}}
\newcommand{\lya}{\ifmmode\mathrm{Ly}\alpha\else{}Ly$\alpha$\fi}
\newcommand{\lyb}{\ifmmode\mathrm{Ly}\beta\else{}Ly$\beta$\fi}
\newcommand{\igm}{\ifmmode\mathrm{IGM}\else{}IGM\fi}
\newcommand{\lae}{\ifmmode\mathrm{LAE}\else{}LAE\fi}
\newcommand{\h}{\ifmmode\mathrm{H}\else{}H\fi}
\newcommand{\hi}{\ifmmode\mathrm{H\,{\scriptscriptstyle I}}\else{}H\,{\scriptsize I}\fi}
\newcommand{\hii}{\ifmmode\mathrm{H\,{\scriptscriptstyle II}}\else{}H\,{\scriptsize II}\fi}
\newcommand{\cmb}{\ifmmode\mathrm{CMB}\else{}CMB\fi}
\newcommand{\qso}{\ifmmode\mathrm{QSO}\else{}QSO\fi}
\newcommand{\eor}{\ifmmode\mathrm{EoR}\else{}EoR\fi}
\newcommand{\heii}{\ifmmode\mathrm{He\,{\scriptscriptstyle II}}\else{}He\,{\scriptsize II}\fi}
\newcommand{\heiii}{\ifmmode\mathrm{He\,{\scriptscriptstyle III}}\else{}He\,{\scriptsize III}\fi}
\newcommand{\ciii}{\ifmmode\mathrm{C\,{\scriptscriptstyle III]}}\else{}C\,{\scriptsize III]}\fi}
\newcommand{\oiii}{\ifmmode\mathrm{O\,{\scriptscriptstyle III}}\else{}O\,{\scriptsize III}\fi}
\newcommand{\aliii}{\ifmmode\mathrm{Al\,{\scriptscriptstyle III}}\else{}Al\,{\scriptsize III}\fi}
\newcommand{\mgii}{\ifmmode\mathrm{Mg\,{\scriptscriptstyle II}}\else{}Mg\,{\scriptsize II}\fi}
\newcommand{\fe}{\ifmmode\mathrm{Fe}\else{}Fe\fi}
\newcommand{\nv}{\ifmmode\mathrm{N\,{\scriptscriptstyle V}}\else{}N\,{\scriptsize V}\fi}
\newcommand{\niv}{\ifmmode\mathrm{N\,{\scriptscriptstyle IV]}}\else{}N\,{\scriptsize IV]}\fi}
\newcommand{\cii}{\ifmmode\mathrm{C\,{\scriptscriptstyle II}}\else{}C\,{\scriptsize II}\fi}
\newcommand{\civ}{\ifmmode\mathrm{C\,{\scriptscriptstyle IV}}\else{}C\,{\scriptsize IV}\fi}
\newcommand{\siv}{\ifmmode\mathrm{Si\,{\scriptscriptstyle IV}}\else{}Si\,{\scriptsize IV}\fi}
\newcommand{\siii}{\ifmmode\mathrm{Si\,{\scriptscriptstyle II}}\else{}Si\,{\scriptsize II}\fi}
\newcommand{\siiii}{\ifmmode\mathrm{Si\,{\scriptscriptstyle III]}}\else{}Si\,{\scriptsize III]}\fi}
\newcommand{\ovi}{\ifmmode\mathrm{O\,{\scriptscriptstyle VI}}\else{}O\,{\scriptsize VI}\fi}
\newcommand{\sioiv}{\ifmmode\mathrm{Si\,{\scriptscriptstyle IV}\,\plus O\,{\scriptscriptstyle IV]}}\else{}Si\,{\scriptsize IV}\,+O\,{\scriptsize IV]}\fi}

\newcommand{\cmmc}{\textsc{\small 21CMMC}}
\newcommand{\cmfst}{\textsc{\small 21CMFAST}}

\pdfoutput=1
\title[MWA astrophysics]{Exploring reionisation and high-$z$ galaxy observables with recent multi-redshift MWA upper limits on the 21-cm signal}
\author[B. Greig et al.] {Bradley~Greig$^{1,2}$\thanks{E-mail:~greigb@unimelb.edu.au}, Cathryn M. Trott$^{2,3}$, Nichole Barry$^{1,2}$, Simon J. Mutch$^{1,2}$, \newauthor Bart Pindor$^{1,2}$, Rachel L. Webster$^{1,2}$, J. Stuart B. Wyithe$^{1,2}$. \\
$^1$School of Physics, University of Melbourne, Parkville, VIC 3010, Australia \\
$^2$ARC Centre of Excellence for All-Sky Astrophysics in 3 Dimensions (ASTRO 3D) \\
$^3$International Centre for Radio Astronomy Research (ICRAR), Curtin University, Bentley, WA, Australia
}

\begin{document}
\maketitle \begin{abstract}
\noindent
We use the latest multi-redshift ($z=6.5-8.7$) upper limits on the 21-cm signal from the Murchison Widefield Array (MWA) to explore astrophysical models which are inconsistent with the data. These upper limits are achieved using 298 h of carefully excised data over four observing seasons. To explore these upper limits in the context of reionisation astrophysics, we use \cmmc{}, a Monte Carlo Markov Chain sampler of full 3D semi-numerical simulations of the cosmic 21-cm signal from \cmfst{}. Further, we connect these disfavoured regions of parameter space to existing observational constraints on the epoch of reionisation such as high-$z$ galaxy ultra-violet (UV) luminosity functions, background UV photoionisation rate, intergalactic medium (IGM) neutral fraction, the electron scattering optical depth and the soft-band X-ray emissivity. We find that the vast majority of astrophysical models disfavoured by the latest MWA limits are already inconsistent with existing observational constraints. These inconsistent models arise from two classes of models: (i) `cold' reionisation and (ii) pure matter density fluctuations (i.e. no reionisation). However, there are a small subsample of models which are consistent. This implies that the existing MWA limits are already beginning to provide unique information to disfavour astrophysical models of reionisation, albeit extremely weakly. Using these latest MWA upper limits on the 21-cm signal, we provide the first limits on the soft-band X-ray emissivity from galaxies at high redshifts. We find $1\sigma$ lower limits of $\epsilon_{{\rm X},0.5-2~{\rm keV}}\gtrsim10^{34.5}$~erg~s$^{-1}$~Mpc$^{-3}$.
Finally, we place limits on the IGM spin temperature, disfavouring values below $\bar{T}_{\rm S}\lesssim$~1.3, 1.4, 1.5, 1.8, 2.1, 2.4~K at $z=6.5, 6.8, 7.1, 7.8, 8.2, 8.7$ at 95 per cent confidence. We infer from this that the IGM must have undergone, at the very least, a small amount of X-ray heating. Note that the limits on both $\epsilon_{{\rm X},0.5-2~{\rm keV}}$ and $\bar{T}_{\rm S}$ are conditional on the IGM neutral fraction.

\end{abstract} 

\section{Introduction}

Prior to the formation of the first stars and galaxies, the Universe is opaque to visible radiation due to the neutral hydrogen fog that pervades. This fog can only be lifted once the cumulative ionising radiation escaping from the stars and galaxies exceed the recombination rate of the neutral hydrogen. This transition is referred to as the Epoch of Reionisation (EoR), and corresponds to the final major baryonic phase change of the Universe. Observing this phase transition is vitally important as it can reveal insights into the properties of the first astrophysical sources. 

Our most promising avenue for detecting the EoR is through observing the 21-cm hyperfine transition of the neutral hydrogen in the intergalactic medium (IGM) \citep[see e.g.][]{Furlanetto:2006p209,Morales:2010p1274,Pritchard:2012p2958,Zaroubi:2013p2976,Barkana:2016}. The radiation from these first stars and galaxies leaves an imprint on the thermal and ionisation state of the IGM, detectable via this 21-cm signal. Importantly, since it is a line transition, the spatial and frequency (redshift) dependence of the 21-cm signal can yield a three dimensional time-lapse of the history of the IGM. From this, we will be able to infer the ultra-violet (UV) and X-ray properties of the astrophysical sources responsible for reionisation.

Unfortunately, observing the cosmic 21-cm signal is extremely difficult, owing to the fact that the signal is five orders of magnitude fainter than the astrophysical foregrounds. Nevertheless this has not deterred numerous experiments from embarking on detecting this elusive signal. These experiments can be broadly classified into two types: (i) global signal experiments which spatially average the signal over the entire sky and (ii) large-scale interferometers sensitive to the spatial fluctuations in the 21-cm signal.

Global signal experiments are conceptually simpler as they typically consist of only a single dipole. Completed or ongoing experiments include, the Experiment to Detect the Global EoR Signature (EDGES; \citealt{Bowman:2010p6724}), the Sonda Cosmol\'{o}gica de las Islas para la Detecci\'{o}n de Hidr\'{o}geno Neutro (SCI-HI; \citealt{Voytek:2014p6741}), the Shaped Antenna measurement of the background RAdio Spectrum (SARAS; \citealt{Patra:2015p6814}), Broadband Instrument for Global HydrOgen ReioNisation Signal (BIGHORNS; \citealt{Sokolowski:2015p6827}), the Large Aperture Experiment to detect the Dark Ages (LEDA; \citealt{Greenhill:2012p6829,Bernardi:2016p6834}), Probing Radio Intensity at high-Z from Marion (PRI$^{\rm Z}$M; \citealt{Philip:2019}) and the Netherlands-China Low-Frequency Explorer (NCLE\footnote{https://www.isispace.nl/projects/ncle-the-netherlands-china-low-frequency-explorer/}).

The first generation of large-scale radio interferometers, the Low-Frequency Array (LOFAR; \citealt{vanHaarlem:2013p200}), the Murchison Wide Field Array (MWA; \citealt{Tingay:2013p2997,Wayth:2018}) and the Precision Array for Probing the Epoch of Reionisation (PAPER; \citealt{Parsons:2010p3000}) have limited sensitivities, requiring long integration times to potentially yield  a low signal-to-noise detection of the cosmic 21-cm signal. These experiments have additionally informed the development of the next generation of significantly larger interferometers; the Hydrogen Epoch of Reionization Array (HERA; \citealt{DeBoer:2017p6740}) and the Square Kilometre Array (SKA; \citealt{Mellema:2013p2975,Koopmans:2015}). With these next generation instruments, not only should we be able to provide high signal-to-noise statistical detections across multiple redshifts, we should also be able to provide the first three-dimensional tomographic images of the EoR.

Other than a reported detection of an absorption feature near $z\approx17$ by EDGES \citep{Bowman:2018}, whose cosmological origins are still heavily questioned \citep[see e.g.][]{Hills:2018,Draine:2018,Bowman:2018b,Bradley:2019,Singh:2019}, we are left only with upper limits on the 21-cm signal. For example, limits on the sky-averaged signal have been obtained with LEDA \citep{Bernardi:2016}, EDGES high-band \citep{Monsalve:2017} and SARAS2 \citep{Singh:2017}. Upper limits on the 21-cm spatial fluctuations, measured through the power spectrum (PS), have been measured at many frequencies (redshifts) and spatial scales throughout the literature, making direct comparisons complicated. The first upper limits were achieved with the Giant Metrewave Radio Telescope (GMRT; \citealt{Paciga:2013}) at $z\approx8.6$. Since then, upper limits have been published by LOFAR at $z=9.6-10.6$ \citep{Patil:2017} and $z=19.8-25.2$ \citep{Gehlot:2019}, PAPER at $z\approx7.5-11$ \citep{Cheng:2018,Kolopanis:2019}, MWA at $z\sim7$ \citep{Dillon:2015,Beardsley:2016,Barry:2019,Li:2019} and by the Owens Valley Radio Observatory Long Wavelength Array (OVRO-LWA; \citealt{Eastwood:2019}) at $z\approx18.4$.

Recently, both LOFAR \citep{Mertens:2020} and the MWA \citep{Trott:2020} published their current best upper limits on the 21-cm PS. For LOFAR this culminated in a best upper limit at $z\approx9.1$ from 141 hours of observations. In the case of the MWA, following careful quality control of the data, deep multi-redshift limits were achieved from 298 h of observations at six separate redshifts spanning $z=6.5-8.7$. While these new limits still remain a few orders of magnitude above fiducial theoretical models, they are aggressive enough to explore extreme models of reionisation. For example, `cold' reionisation, where the 21-cm PS amplitude can be in excess of $\Delta^{2}_{21}\gtrsim10^4~{\rm mK}^{2}$ due to large temperature contrasts between the neutral and ionised IGM \citep[e.g.][]{Mesinger:2014p244,Parsons:2014p781}. This occurs when the neutral IGM undergoes little to no heating and adiabatically cools faster than the Cosmic Microwave Background (CMB) temperature with the expansion of the Universe.

In this work, we shall focus explicitly on exploring the astrophysical models inconsistent with the MWA multi-redshift limits from \citet{Trott:2020}. Similar analyses have already been performed for LOFAR in the context of general IGM properties such as the neutral fraction and spin temperature \citep{Ghara:2020}, the excess radio background \citep{Mondal:2020} and high-$z$ galaxy and reionisation observables \citet{Greig:2020}. Here, we follow the same approach as \citet{Greig:2020}. 

Our exploration takes advantage of \cmmc{}\footnote{https://github.com/BradGreig/21CMMC} \citep{Greig:2015p3675,Greig:2017p8496,Greig:2018,Park:2019}, a Monte-Carlo Markov Chain (MCMC) sampler of the semi-numerical reionisation code \cmfst{}\footnote{https://github.com/andreimesinger/21cmFAST}\citep{Mesinger:2007p122,Mesinger:2011p1123}. \cmmc{} forward models the full 3D cosmic 21-cm signal provided by \cmfst{} in a fully Bayesian framework allowing us to compare against observations of the first billion years of the Universe. In particular, we adopt the \citet{Park:2019} galaxy model parameterisation allowing direct comparison against high-$z$ galaxy ultra-violet (UV) luminosity functions (LFs).

The outline of this work is as follows. In Section~\ref{sec:Method}, we summarise the \cmfst{} astrophysical model before outlining the \cmmc{} setup in Section~\ref{sec:setup}. In Section~\ref{sec:results} we discuss our main results before providing our conclusions in Section~\ref{sec:Conclusion}. Unless otherwise stated, all quantities are in co-moving units with the following adopted cosmological parameters:  ($\Omega_\Lambda$, $\Omega_{\rm M}$, $\Omega_b$, $n$, $\sigma_8$, $H_0$) = (0.69, 0.31, 0.048, 0.97, 0.81, 68 km s$^{-1}$ Mpc$^{-1}$), consistent with recent results from the Planck mission \citep{PlanckCollaboration:2016p7780}.

\section{Modelling the 21-cm signal} \label{sec:Method}

The cosmic 21-cm signal is modelled using the semi-numerical simulation code, \cmfst{} \citep{Mesinger:2007p122,Mesinger:2011p1123}. In particular, we use the \citet{Park:2019} astrophysical parameterisation, which allows the star-formation rate and ionising escape fraction to depend on the mass of the host dark matter halo. This, following some simple conversions, enables \cmfst{} to produce UV LFs that are able to be compared against observed high-$z$ galaxy LFs. Additionally, we also include a recipe for an on-the-fly ionising photon non-conservation correction (Park et al., in prep) to account for the fact that 3D excursion set approaches that track ionisations are not photon conserving \citep[e.g.][]{McQuinn:2005,Zahn:2007,Paranjape:2014,Paranjape:2016,Hassan:2017,Choudhury:2018,Hutter:2018,Molaro:2019}. Below, we shall outline the important aspects of \cmfst{} which lead to the modelling of the 21-cm signal, and defer the reader to the aforementioned publications for further details.

\subsection{Galaxy UV Properties}

The typical stellar mass of a galaxy, $M_{\ast}$, is assumed to be directly related to its host halo mass, $M_{\rm h}$ \citep[e.g.][]{Kuhlen:2012p1506,Dayal:2014b,Behroozi:2015p1,Mitra:2015,Mutch:2016,Ocvirk:2016,Sun:2016p8225,Yue:2016,Hutter:2020}:
\begin{eqnarray} \label{}
M_{\ast}(M_{\rm h}) = f_{\ast}\left(\frac{\Omega_{\rm b}}{\Omega_{\rm m}}\right)M_{\rm h},
\end{eqnarray}
where $f_{\ast}$ is the fraction of galactic gas in stars given by:
\begin{eqnarray} \label{}
f_{\ast} = f_{\ast, 10}\left(\frac{M_{\rm h}}{10^{10}\,M_{\odot}}\right)^{\alpha_{\ast}},
\end{eqnarray}
with $f_{\ast, 10}$ being its normalisation and a power-law\footnote{A power-law dependence between $M_{\ast}$ and $M_{\rm h}$ at $z\gtrsim5$ is consistent with the mean behaviour of both semi-analytic model predictions \citep[e.g][]{Mutch:2016,Yung:2019,Hutter:2020} and semi-empirical fits to observations \citep[e.g.][]{Harikane:2016,Tacchella:2018,Behroozi:2019}.} index, $\alpha_{\ast}$. 

The star-formation rate (SFR) is then estimated by dividing the stellar mass by a characteristic time-scale,
\begin{eqnarray} \label{eq:Mdt}
\dot{M}_{\ast}(M_{\rm h},z) = \frac{M_{\ast}}{t_{\ast}H^{-1}(z)},
\end{eqnarray}
where $t_{\ast}$ is a free parameter allowed to vary between zero and unity and $H^{-1}(z)$ is the Hubble time.

A galaxy's UV ionising escape fraction, $f_{\rm esc}$, is equivalently allowed to vary with halo mass,
\begin{eqnarray} \label{}
f_{\rm esc} = f_{\rm esc, 10}\left(\frac{M_{\rm h}}{10^{10}\,M_{\odot}}\right)^{\alpha_{\rm esc}},
\end{eqnarray}
with normalisation set by $f_{\rm esc, 10}$ and a power-law index, $\alpha_{\rm esc}$.

Finally, to account for inefficient cooling and/or feedback processes which can prevent small mass halos from hosting active, star-forming galaxies, a duty-cycle is included to suppress their contribution:
\begin{eqnarray} \label{eq:duty}
f_{\rm duty} = {\rm exp}\left(-\frac{M_{\rm turn}}{M_{\rm h}}\right).
\end{eqnarray}
This results in a fraction, $(1 - f_{\rm duty})$, of host haloes which do not host star-forming galaxies, whose suppression scale is controlled by $M_{\rm turn}$ \citep[e.g.][]{Shapiro:1994,Giroux:1994,Hui:1997,Barkana:2001p1634,Springel:2003p2176,Mesinger:2008,Okamoto:2008p2183,Sobacchi:2013p2189,Sobacchi:2013p2190}.

In summary, we have six free parameters, $f_{\ast, 10}$, $f_{\rm esc, 10}$, $\alpha_{\ast}$, $\alpha_{\rm esc}$, $M_{\rm turn}$ and $t_{\ast}$, which describe the galaxy UV properties. 

\subsection{Galaxy X-ray Properties}

In the early Universe, X-rays escaping from the first galaxies, likely from stellar remnants, are thought to be responsible for the heating of the IGM. X-ray heating in \cmfst{} is modelled by calculating the cell-by-cell angle-averaged specific X-ray intensity, $J(\boldsymbol{x}, E, z)$, (in erg s$^{-1}$ keV$^{-1}$ cm$^{-2}$ sr$^{-1}$). This is obtained by integrating the co-moving X-ray specific emissivity, $\epsilon_{\rm X}(\boldsymbol{x}, E_e, z')$ back along the light-cone:
\begin{equation} \label{eq:Jave}
J(\boldsymbol{x}, E, z) = \frac{(1+z)^3}{4\pi} \int_{z}^{\infty} dz' \frac{c dt}{dz'} \epsilon_{\rm X}  e^{-\tau},
\end{equation}
where $e^{-\tau}$ accounts for attenuation by the IGM. In the emitted frame, $E_{\rm e} = E(1 + z')/(1 + z)$, the co-moving specific emissivity is,
\begin{equation} \label{eq:emissivity}
\epsilon_{\rm X}(\boldsymbol{x}, E_{\rm e}, z') = \frac{L_{\rm X}}{\rm SFR} \left[ (1+\bar{\delta}_{\rm nl}) \int^{\infty}_{0}{\rm d}M_{\rm h} \frac{{\rm d}n}{{\rm d}M_{\rm h}}f_{\rm duty} \dot{M}_{\ast}\right],
\end{equation}
where $\bar{\delta}_{\rm nl}$ is the mean, non-linear density in a shell around $(\boldsymbol{x}, z)$ and the quantity in square brackets is the SFR density along the light-cone. 

This expression is normalised by the specific X-ray luminosity per unit star formation escaping the host galaxies, $L_{\rm X}/{\rm SFR}$ (erg s$^{-1}$ keV$^{-1}$ $M^{-1}_{\odot}$ yr). The X-ray luminosity is assumed to be a power-law with respect to photon energy, $L_{\rm X} \propto E^{- \alpha_{\rm X}}$, which is attenuated below a threshold energy, $E_0$, to account for absorption of low energy X-rays by a neutral interstellar medium within the host galaxy. Finally, this specific luminosity is then normalised to an integrated soft-band ($<2$~keV) luminosity per SFR (in erg s$^{-1}$ $M^{-1}_{\odot}$ yr), which is a free parameter in the model:
\begin{equation} \label{eq:normL}
  L_{{\rm X}<2\,{\rm keV}}/{\rm SFR} = \int^{2\,{\rm keV}}_{E_{0}} dE_e ~ L_{\rm X}/{\rm SFR} ~.
\end{equation}
A limit of $2\,{\rm keV}$ corresponds to a mean-free path of the order of the Hubble length at high redshifts, implying harder X-ray photons do not contribute to IGM heating \citep[e.g.][]{McQuinn:2012p3773}.

In summary, there are three free model parameters describing the X-ray properties of the first galaxies,  $L_{{\rm X}<2\,{\rm keV}}/{\rm SFR}$ , $E_{0}$ and $\alpha_{\rm X}$.

\subsection{Ionisation and Thermal State of the IGM}

At any redshift within \cmfst{} the evolved IGM density and velocity fields are obtained following second-order Lagrange perturbation theory \citep[e.g][]{Scoccimarro:1998p7939} from an initial high resolution linear density field. The ionisation field is then determined from the evolved density field by the excursion-set approach \citep{Furlanetto:2004p123}, whereby the balance between the cumulative number of ionising photons and the number of neutral hydrogen atoms plus cumulative recombinations are tracked within spheres of decreasing radii. A cell is considered ionised when,
\begin{eqnarray} \label{eq:ioncrit}
n_{\rm ion}(\boldsymbol{x}, z | R, \delta_{R}) \geq (1 + \bar{n}_{\rm rec})(1-\bar{x}_{e}),
\end{eqnarray}
where $\bar{n}_{\rm rec}$ is the cumulative number of recombinations \citep[e.g.][]{Sobacchi:2014p1157} and $(1-\bar{x}_{e})$, corresponds to ionisations by X-rays, which are expected to contribute at the $\sim 10$ per cent level \citep[e.g.][]{Ricotti:2004p7145,Mesinger:2013p1835,Madau:2017,Ross:2017,Eide:2018}. This first term, $n_{\rm ion}$, is the cumulative number of IGM ionising photons per baryon inside a spherical region of size, $R$ and corresponding overdensity, $\delta_{R}$,
\begin{eqnarray} \label{eq:ioncrit2}
n_{\rm ion} = \bar{\rho}^{-1}_b\int^{\infty}_{0}{\rm d}M_{\rm h} \frac{{\rm d}n(M_{h}, z | R, \delta_{R})}{{\rm d}M_{\rm h}}f_{\rm duty} \dot{M}_{\ast}f_{\rm esc}N_{\gamma/b},
\end{eqnarray}
where $\bar{\rho}_b$ is the mean baryon density and $N_{\gamma/b}$ is the number of ionising photons per stellar baryon\footnote{We assume $N_{\gamma/b}=5000$, corresponding to a Salpeter initial mass function \citep{Salpeter:1955}; however note that this is highly degenerate with the fraction of galactic gas in stars, $f_{\ast}$.}.

\subsection{Ionising Photon Non-Conservation Correction}

Excursion-set approaches for tracking ionisations in three dimensions, like that employed by \cmfst{}, do not conserve ionising photons. This is driven by the surplus of ionising photons remaining within a cell after exceeding the ionisation criteria (Eq.~\ref{eq:ioncrit}) that are not propagated further into the IGM. In effect, this results in an effective bias on the ionising escape fraction, $f_{\rm esc}$. This behaviour has been studied extensively in the literature and has been shown to result in a loss of $\sim10-20$ per cent of the ionising photons \citep[e.g.][]{McQuinn:2005,Zahn:2007,Paranjape:2014,Paranjape:2016,Hassan:2017,Choudhury:2018,Hutter:2018,Molaro:2019}. 

Recently, explicit photon conserving algorithms for semi-numerical simulations have been introduced by \citet{Choudhury:2018} and \citet{Molaro:2019}. However, despite being orders of magnitude faster than full radiative-transfer simulations they remain 
too slow when forward modelling the 21-cm signal in the high-dimensional parameter spaces required to characterise the ionising, soft UV, and X-ray properties of the first galaxies.

Alternatively, Park et al., in prep, introduce an approximate correction to the effective bias on $f_{\rm esc}$ by analytically solving for the evolution of the ionisation fraction given a source model, assuming no correlations between the sources and sinks\footnote{This assumption only breaks down for the final $\sim10$ per cent of the EoR \citep{Sobacchi:2014p1157}.}. By comparing this analytic expression against a calibration curve (generated from \cmfst{} including only ionisations) the delay in the resultant reionisation history (owing to the loss of photons) can be corrected for by modifying the redshift at which ionisations are determined. For the duration of the EoR, this correction results in a shift in redshift of $\Delta z\sim0.3 \pm 0.1$.

\subsection{21-cm Brightness Temperature}

Finally, the observed 21-cm signal is measured as a brightness temperature contrast relative to the CMB temperature, $T_{\rm CMB}$ \citep[e.g.][]{Furlanetto:2006p209}:
\begin{eqnarray} \label{eq:21cmTb}
\delta T_{\rm b}(\nu) &=& \frac{T_{\rm S} - T_{\rm CMB}(z)}{1+z}\left(1 - {\rm e}^{-\tau_{\nu_{0}}}\right)~{\rm mK},
\end{eqnarray}
where $\tau_{\nu_{0}}$ is the optical depth of the 21-cm line,
\begin{eqnarray}
\tau_{\nu_{0}} &\propto& (1+\delta_{\rm nl})(1+z)^{3/2}\frac{x_{\hi{}}}{T_{\rm S}}\left(\frac{H}{{\rm d}v_{\rm r}/{\rm d}r+H}\right).
\end{eqnarray}
Here, $x_{\hi{}}$ corresponds to the neutral hydrogen fraction, $\delta_{\rm nl} \equiv \rho/\bar{\rho} - 1$ is the gas density, $H(z)$ is the Hubble parameter, ${\rm d}v_{\rm r}/{\rm d}r$ is the line-of-sight component of the velocity gradient and $T_{\rm S}$ is the gas spin temperature. All quantities, obtained following the method described in the previous sections are evaluated at redshift $z = \nu_{0}/\nu - 1$, where $\nu_{0}$ is the 21-cm frequency and we drop the spatial dependence for brevity. Redshift space distortions along the line-of-sight are additionally included as outlined in \citet{Mao:2012p7838,Jensen:2013p1389,Greig:2018}.

\section{\cmmc{} setup} \label{sec:setup}

\cmmc{} \citep{Greig:2015p3675,Greig:2017p8496,Greig:2018,Park:2019} is the publicly available, massively parallel MCMC sampler of the 3D semi-numerical reionisation simulation code \cmfst{} \citep{Mesinger:2007p122,Mesinger:2011p1123}. It is based on the publicly available \textsc{\small Python} module \textsc{\small CosmoHammer} \citep{Akeret:2012p842}. MCMC sampling is performed using the \textsc{\small Emcee} \textsc{\small Python} module \citep{ForemanMackey:2013p823}, which is an affine invariant ensemble sampler from \citet{Goodman:2010p843}. For each proposed parameter set in the MCMC, \cmmc{} performs an independent 3D realisation of the 21-cm signal, allowing any quantity to be sampled (e.g. the 21-cm PS). Below, we outline the astrophysical parameter set to be explored within this work (Section~\ref{sec:fiducial}), the interpretation of the MWA upper limits (Section~\ref{sec:limits}) and the \cmmc{} setup (Section~\ref{sec:sims}).

\subsection{Astrophysical parameter set} \label{sec:fiducial}

For the adopted model described in Section~\ref{sec:Method}, we have nine astrophysical parameters. Below, we summarise each parameter as well as the corresponding assumed parameter ranges (which we have adopted based on previous works, e.g. \citealt{Greig:2017,Park:2019}). Additionally, we summarise these parameter ranges in the top row of Table~\ref{tab:Results}. Throughout, we assume flat priors on each astrophysical parameters.

\begin{itemize}
\item[(i)]$f_{\ast, 10}$: the fraction of galactic gas in stars evaluated at a halo mass of 10$^{10}~M_{\odot}$. The log quantity is allowed to vary as, ${\rm log}_{10}(f_{\ast, 10}) \in [-3,0]$.
\item[(ii)]$\alpha_{\ast}$: the power-law index for the halo mass dependent star-formation. This is varied between, $\alpha_{\ast} \in [-0.5,1]$.
\item[(iii)]$f_{\rm esc, 10}$: the ionising UV escape fraction evaluated at a halo mass of 10$^{10}~M_{\odot}$. The log of this quantity varies between, ${\rm log}_{10}(f_{\rm esc, 10}) \in [-3, 0]$.
\item[(iv)]$\alpha_{\rm esc}$: the power-law index for halo mass dependent ionising UV escape fraction. This is varied between, $\alpha_{\rm esc} \in [-1,0.5]$.
\item[(v)]$t_{\ast}$: the star-formation time scale as a fraction of the Hubble time, which is varied in the range, $t_{\ast} \in (0,1]$.
\item[(vi)]$M_{\rm turn}$: the characteristic halo mass below which the abundance of active star-forming galaxies are exponentially suppressed by a duty cycle (see Equation~\ref{eq:duty}). We allow this to vary between, ${\rm log}_{10}(M_{\rm turn}) \in[8,10]$.
\item[(vii)]$E_{0}$: the minimum energy threshold above which X-ray photons can escape their host galaxy. We allow this to vary between, $E_{0} \in [0.2,1.5]$~keV, which corresponds of an integrated column density of, ${\rm log_{10}}(N_{\hi{}}/{\rm cm}^{2}) \in [19.3,23.0]$.
\item[(viii)]$L_{{\rm X}<2\,{\rm keV}}/{\rm SFR}$: the soft-band X-ray luminosity per unit star-formation from the $E_{0} - 2$~keV energy band. We vary this between, ${\rm log_{10}}(L_{{\rm X}<2\,{\rm keV}}/{\rm SFR}) \in [30, 42]$. This lower bound is considerably lower than what is typically adopted \citep[i.e.][]{Park:2019} in-order to explore extreme `cold' reionisation scenarios.
\item[(iv)]$\alpha_{\rm X}$: the power-law index of the X-ray source spectral energy distribution (SED), which we allow to vary between $\alpha_{\rm X} \in [-1,3]$.
\end{itemize}

\subsection{The latest MWA 21-cm upper limits} \label{sec:limits}

Recently, \citet{Trott:2020} published deep, multi-redshift upper limits on the 21-cm signal from the EoR using the MWA. These limits are obtained at $k = 0.07 - 3.0~h~{\rm Mpc}^{-1}$ across six redshift bins from $z=6.5-8.7$ using 298~h of carefully selected clean observations over four observing seasons. At $z=6.5$, these correspond to the lowest yet available upper limits on the reionisation epoch.

Unfortunately, these upper limits are still too large to begin to provide statistical constraints on astrophysical models of reionisation. However, we can instead use these upper limits to explore astrophysical models that {\it exceed} (i.e. are disfavoured by) the existing observational limits placed by the MWA. This follows the same approach to that applied to the recent upper limits on the 21-cm signal at $z\approx9.1$ achieved by LOFAR (\citealt{Ghara:2020,Greig:2020,Mondal:2020}).

In order to explore astrophysical models disfavoured by the MWA, we construct a likelihood function of the following form:
\begin{eqnarray}
\mathcal{L}(\theta) \propto \prod^{n}_{i,j} \frac{\int^{(1 + \sigma)\Delta^{2}_{21,{\rm mod.}}}_{(1 - \sigma)\Delta^{2}_{21,{\rm mod.}}} p_{\rm ex.}(\Delta^{2}_{21,{\rm mod.}}(k_{i},z_{j},\theta)) {\rm d}\Delta^{2}_{21,{\rm mod.}}}{\int^{(1 + \sigma)\Delta^{2}_{21,{\rm mod.}}}_{(1 - \sigma)\Delta^{2}_{21,{\rm mod.}}} {\rm d}\Delta^{2}_{21,{\rm mod.}}}
\end{eqnarray}
Here, $\theta$ corresponds to the astrophysical parameter set (i.e. model), $\Delta^{2}_{21, {\rm mod.}}(k_{i},z_{j},\theta)$ is the model 21-cm PS (from \cmfst{}) as a function of $k$ and $z$ and $p_{\rm ex.}(\Delta^{2}_{21,{\rm mod.}}(k_{i},z_{j},\theta))$ is the probability of the model 21-cm PS exceeding the MWA upper-limit. The integral over the range $\pm\sigma$ accounts for uncertainties on the amplitude of the modelled 21-cm PS.

The probability of the 21-cm PS to be in excess of the upper limits is obtained from the true, measured probability density functions (PDFs) of the 21-cm PS amplitude used to construct the observed upper limits from \citet{Trott:2020}. In effect, $p_{\rm ex.}(\Delta^{2}_{21,{\rm mod.}}(k_{i},z_{j},\theta))$ goes to zero for decreasing 21-cm PS amplitudes below the MWA upper limits and approaches unity for model 21-cm PS amplitudes far in excess of the upper-limit. 

In this work, we combine the data from all six redshift bins ($z=6.5, 6.8, 7.1, 7.8, 8.2$ and 8.7) with the first four Fourier bins, $k=0.14, 0.21, 0.28$ and 0.35 $h~$Mpc$^{-1}$. Note that for smaller scales (i.e. larger $k$), the upper limits (and associated errors) become sufficiently large that they provide very little additional information (see e.g. Figure~\ref{fig:21cmPS}).

\subsection{Simulations} \label{sec:sims}

Now, with our likelihood described as above, we outline the simulation setup adopted for \cmmc{}. In order to sufficiently sample the upper-limit at the largest spatial scale ($k=0.14~h$~Mpc$^{-1}$), we perform 3D realisations of the 21-cm signal in a volume with length 250~Mpc and 128 voxels per side-length. For our total uncertainty, $\sigma$, on the modelled 21-cm PS we sum in quadrature a conservative 20 per cent multiplicative modelling uncertainty\footnote{This modelling uncertainty is motivated by approximations adopted in semi-numerical simulations relative to radiative-transfer simulations \citep[e.g.][]{Zahn:2011p1171,Ghara:2018,Hutter:2018}.} on the sampled 21-cm PS with the estimated sample variance from our simulation setup. Note that for this simulation setup, the estimated sample variance on the 21-cm PS at $k=0.14~h$~Mpc$^{-1}$ is $\lesssim10$ per cent.

\section{Results} \label{sec:results}

\subsection{Disfavoured 21-cm PS} \label{sec:21cmPS}

\begin{figure*} 
	\begin{center}
	  \includegraphics[trim = 2cm 1.4cm 0cm 1.5cm, scale = 0.31]{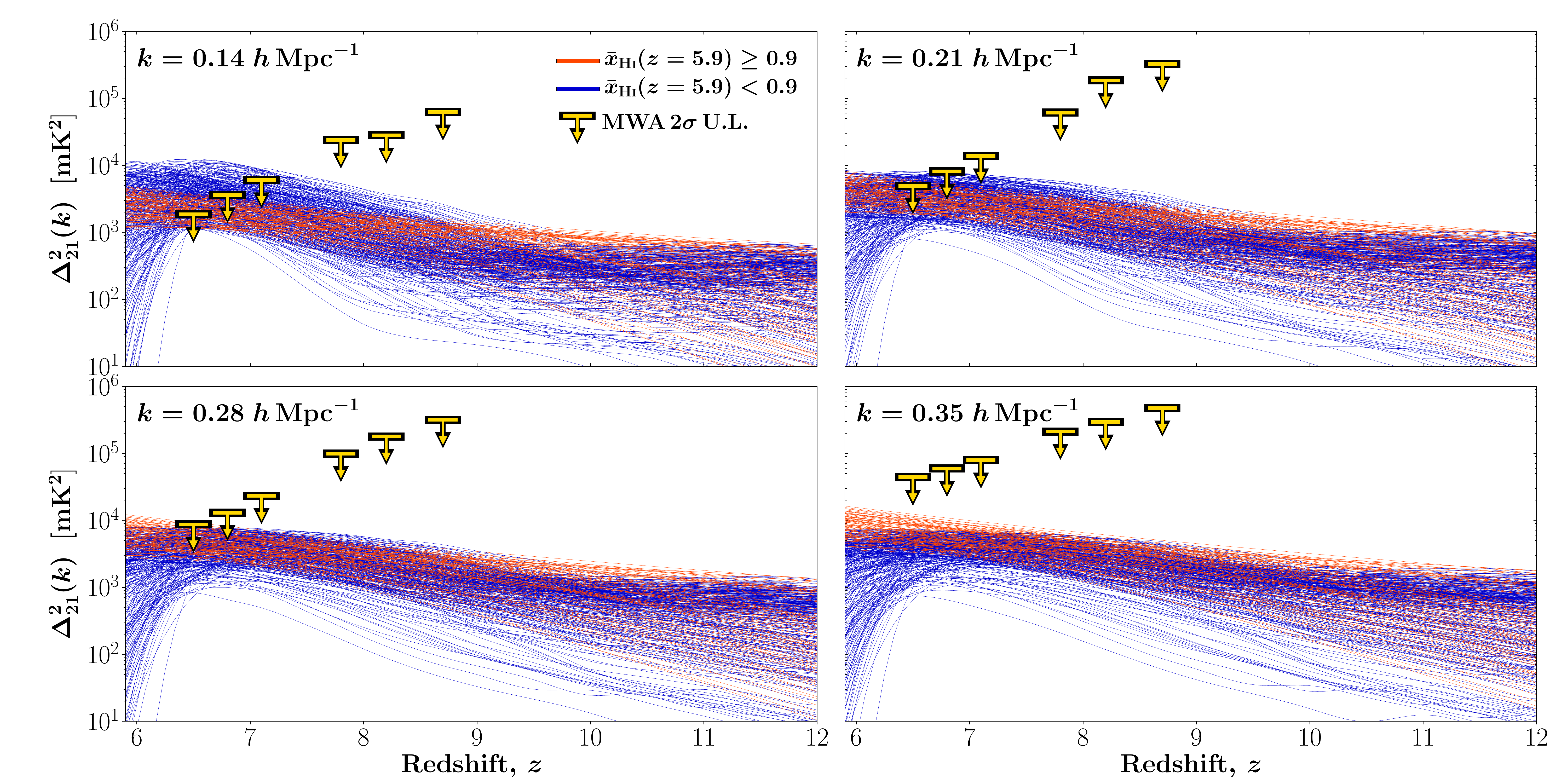}
	\end{center}
\caption[]{A random sample of 500 21-cm PS as a function of redshift disfavoured by the latest MWA upper limits \citep{Trott:2020}. The four panels correspond to the first four Fourier modes ($k = 0.14, 0.21, 0.28, 0.35~h~{\rm Mpc^{-1}}$). The upper limits (yellow) correspond to the 2$\sigma$ MWA upper limits \citep{Trott:2020}, whereas the blue (red) curves correspond to the 21-cm PS colour coded by the resultant neutral fraction at $z=5.9$.}
\label{fig:21cmPS}
\end{figure*}

Before exploring the implications of the latest MWA upper limits \citep{Trott:2020} on the individual astrophysical parameters describing reionisation, it is first illustrative to explore the 21-cm PS which are disfavoured by the observational limits. In Figure~\ref{fig:21cmPS} we show the evolution in the 21-cm PS as a function of redshift for the first four Fourier modes ($k = 0.14, 0.21, 0.28, 0.35~h~{\rm Mpc^{-1}}$). The thin blue and red curves correspond to 500 randomly sampled 21-cm PS which are separated by the resultant neutral fraction at $z=5.9$ in order to distinguish between two types of models.

Typically, when presenting the 21-cm PS evolution as a function of redshift, one expects to observe three distinctive peaks at large-scales in the PS amplitude \cite[e.g.][]{Pritchard:2007,Baek:2010p6357,Mesinger:2011p1123}. These are, in order of increasing redshift, driven by (i) reionisation, (ii) X-ray heating and (iii) Wouthuysen--Field (WF) coupling. In the case of these extreme `cold' reionisation models (i.e. absence of X-ray heating), we expect no secondary peak, with a single large amplitude peak during reionisation along with a WF-coupling peak at higher redshifts.

This reionisation peak is clearly exhibited by the blue curves in Figure~\ref{fig:21cmPS}, which corresponds to IGM neutral fractions below 90 per cent at $z = 5.9$. Here, the 21-cm PS amplitude peaks at $z=6-7$ before rapidly dropping in amplitude as reionisation completes. These models are indicative of these `cold' reionisation models, with the fluctuations being driven by the patchy nature of reionisation and the amplitude as a result of the large temperature contrasts in the cold, neutral IGM. These astrophysical models clearly exceed the recent $2\sigma$ upper limits presented by the MWA \citep{Trott:2020}, highlighting that the MWA can already begin to rule out such extreme models.

Note however that the 21-cm PS can only exceed the latest MWA upper limits at the largest scale ($k=0.14~h~{\rm Mpc}^{-1}$). For larger $k$ (smaller scales) the MWA upper limits increase significantly to render all models consistent with the data. Thus, almost all of the constraining power arises from the redshift evolution of the large-scale power. This highlights that, in order to produce the largest gains in ruling out regions of astrophysical parameter space, in the short term observational efforts should focus on improving the 21-cm PS upper limits at large scales, where this `bump' in the redshift evolution occurs as a result of reionisation. 

For the second type of models, corresponding to an IGM neutral fraction above 90 per cent at $z = 5.9$ (denoted by the red curves in Figure~\ref{fig:21cmPS}), we observe no distinctive features. The 21-cm PS amplitude exhibits a smooth increase in amplitude with decreasing redshift. Again, the large amplitudes are driven by the large temperature contrasts in the cold, neutral IGM, however, the high neutral fractions (as indicated by the large IGM neutral fraction at $z=5.9$) are indicative of models where reionisation has yet to commence. Thus, these smooth, featureless 21-cm PS are simply driven by fluctuations in the underlying matter density field. This is the first time astrophysical models driven by pure matter density fluctuations have been found to be disfavoured by limits from the 21-cm signal. 

\subsection{Disfavoured astrophysical parameters} \label{sec:astro}

\begin{figure*} 
	\begin{center}
	  \includegraphics[trim = 0.6cm 1.4cm 0cm 0cm, scale = 0.5]{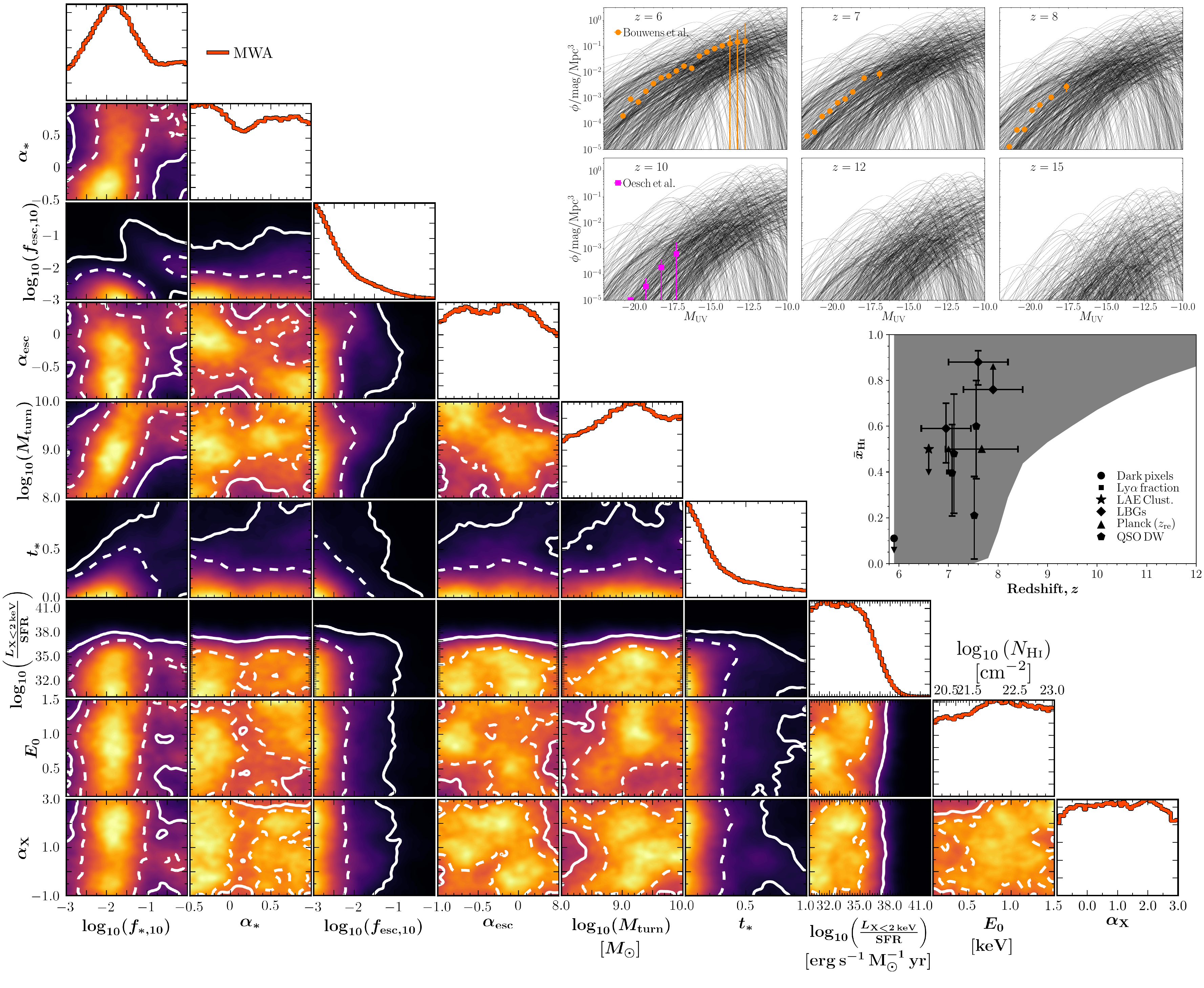}
	\end{center}
\caption[]{Marginalised one and two dimensional posterior distributions for the nine astrophysical parameters which are disfavoured by the multi-frequency upper limits on the 21-cm PS from the MWA \citep{Trott:2020}. White dashed (solid) contours correspond to the 68th (95th) percentiles. In the top right panels, we provide 500 randomly sampled LFs which are drawn from the posterior of astrophysical models that exceed at least one of the MWA upper limits and are compared against existing constraints at $z=6-8$ \citep{Bouwens:2015,Bouwens:2017} and $z=10$ \citep{Oesch:2018}. In the middle right panel, we compare the bounds on the reionisation histories disfavoured by MWA against all current observational constraints on the IGM neutral fraction (see text for further details).}
\label{fig:Corner}
\end{figure*}

\begin{table*}
\tiny
\begin{tabular}{@{}lccccccccc}
\hline
  & ${\rm log_{10}}(f_{\ast,10})$ & $\alpha_{\ast}$ & ${\rm log_{10}}(f_{\rm esc,10})$ & $\alpha_{\rm esc}$ & ${\rm log_{10}}(M_{\rm turn})$ & $t_{\ast}$ & ${\rm log_{10}}\left(\frac{L_{{\rm X}<2{\rm keV}}}{\rm SFR}\right)$ & $E_0$  & $\alpha_{\rm X} $ \\
               &  &  &  &  & $[{\rm M_{\sun}}]$ & & $[{\rm erg\,s^{-1}\,M_{\sun}^{-1}\,yr}]$ &  $[{\rm keV}]$ \\
\hline
\vspace{0.8mm}
Prior ranges & [-3.0, 0.0] & [-0.5,1.0] & [-3.0, 0.0] & [-1.0,0.5] & [8.0,10.0] & (0.0,1.0] & [30.0,42.0] & [0.2,1.5] & [-1.0,3.0] \\
\hline
\vspace{0.8mm}
68th percentile limits & [-2.40, -0.75] &  [-0.5, 0.75]  & [-3.0, -1.75]  &  [-1.0, 0.21]  &  [8.40, 10.0]  &  (0.0, 0.55]  &   [30.0, 35.9]   & [0.35,1.5] & [-1.0,3.0]  \\
95th percentile limits & [-2.88, -0.11] &  [-0.5, 0.96]  & [-3.0, -0.80]  &   [-1.0, 0.45] &  [8.06, 10.0] &    (0.0, 0.91]  &   [30.0, 37.8]   &  [0.14,1.5] & [-1.0,3.0]\\
\hline
\end{tabular}
\caption{Summary of the 68th and 95th percentile limits on the disfavoured regions on the nine astrophysical parameters included in \cmmc{} using the $z=6.5-8.7$ upper limits from the MWA \citep{Trott:2020}.}
\label{tab:Results}
\end{table*} 

In Figure~\ref{fig:Corner}, we present the marginalised one and two dimensional posterior distributions recovered from \cmmc{} for the nine astrophysical parameters disfavoured by the multi-frequency upper limits on the 21-cm PS from the MWA \citep{Trott:2020}. It is important to note that these posteriors are obtained using only the MWA upper limits, with no other existing observation constraints used. Dashed (solid) white contours on the two dimensional posteriors correspond to the 68th (95th) percentile limits. The one dimensional marginalised 68th and 95th percentile limits on each parameter are summarised in Table~\ref{tab:Results}.

At this point it is important to remember that these marginalised posteriors only contain information about the astrophysical models which can exceed the recent MWA upper limits on the 21-cm signal. These marginalised posteriors do not imply that these regions of astrophysical parameter space are ruled out at any statistical significance. Rather, it is indicative of the types of extreme astrophysical models that observational experiments such as the MWA are already starting to disfavour\footnote{These extreme models occupy only a very small sub-volume of the total allowed parameter volume, thus they are not significant enough to result in actual constrains once we marginalise over the full parameter space}. Further, it is a demonstration of the necessity of forward modelling tools such as \cmmc{} for exploring and interpreting observational data. Nevertheless, below we discuss some of the implications on the EoR that can be made from the recent MWA upper limits.

Additionally, it is also worth noting that the results presented in this work are specific to the underlying astrophysical model assumptions. For example, the astrophysical parameterisation used in \cmfst{} assumes only a single ionising source population and further, ignores any explicit redshift dependence on the escape fraction or stellar mass. However, there is currently no evidence for a more complex parameterisation as the simple \citet{Park:2019} model has been shown to be consistent with galaxy UV LFs, semi-numerical SAMs and hydrodynamical simulations.

The main observation we can take from Figure~\ref{fig:Corner} are the limits on the soft-band X-ray luminosity, $L_{{\rm X}<2\,{\rm keV}}/{\rm SFR}$. The primary role of this X-ray luminosity is to control the amount of heating the IGM incurs between the dark ages and the EoR as a result the thermal energy deposited by the X-rays. Low values of this quantity are responsible for producing both: (i) the extreme `cold' reionisation scenarios represented by the blue curves in Figure~\ref{fig:21cmPS} and (ii) the large amplitude 21-cm PS due only to the matter density fluctuations (red curves in Figure~\ref{fig:21cmPS}). With these latest MWA upper limits we recover, at 95 per cent confidence, disfavoured limits of ${\rm log_{10}}\left(L_{{\rm X}<2\,{\rm keV}}/{\rm SFR}\right) \lesssim 37.8$, implying the MWA disfavours models with lower soft-band X-ray luminosities. Presently, these disfavoured limits sit well below our current expectations from observations of analogue low-redshift star-forming galaxies \citep{Mineo:2012p6282}, stacked {\it Chandra} observations \citep{Lehmer:2016p7810} and predictions at high-redshift from population synthesis models \citep{Fragos:2013p6529} (see Section~\ref{sec:Xrays} for more details). In contrast, for the remaining two galaxy X-ray properties, $E_0$ and $\alpha_{\rm X}$, we recover no meaningful limits, owing to the fact we are disfavouring low amplitude soft-band X-ray luminosities. In the absence of X-ray heating, the shape of the X-ray SED does not matter.

Unfortunately, as we are only exploring models in excess of the 21-cm upper limits, it is not fair to perform direct comparisons between these results and those presented by LOFAR at $z\approx9.1$. The primary reason for this is that this approach is sensitive {\it only} to models that maximise the 21-cm PS (in order to exceed the upper limits on the 21-cm signal). In this work, the MWA upper limits are most sensitive to reionisation models that produce 21-cm PS that peak at $z<7.5$ (see Figure~\ref{fig:21cmPS}), whereas the current LOFAR limits are sensitive to only a single redshift at $z\approx9.1$. Further, the existing LOFAR limits are more sensitive to the larger spatial scales (i.e. $k=0.075~h$~Mpc$^{-1}$). Thus the disfavoured regions of each experiment are sensitive to different reionisation models. Nevertheless, one could in principal combine the recent limits from both the MWA and LOFAR, to improve our understanding of these disfavoured regions, however we postpone that to future work.

In terms of the galaxy UV properties, the strongest disfavoured limits are on the fraction of galactic gas in stars, $f_{\ast,10}$, the escape fraction, $f_{\rm esc,10}$, and the SFR time-scale, $t_{\ast}$. Again, these limits are driven by seeking to maximise the 21-cm PS amplitude between $z=6.5-8.7$ (the reionisation `peak' at $z=6-7$ in Figure~\ref{fig:21cmPS}). Qualitatively speaking, this occurs when reionisation is roughly close to its midpoint (i.e. $\bar{x}_{\hi{}} \sim 0.5$; \citealt{Mellema:2006,Lidz:2008p1744})\footnote{However, note that this dependence between the peak of the 21-cm PS and $\bar{x}_{\hi{}}$ is strongly model dependent.}. For example, both $f_{\ast, 10}$ and $f_{\rm esc,10}$, which are highly-degenerate in the absence of constraints from UV LFs \citep[e.g.][]{Park:2019}, control the timing of reionisation. Thus, limits on these quantities are driven to very low values (minimising the number of ionising photons produced), pushing reionisation to occur at lower redshifts (i.e. $z=6-7$). However, rather than producing a limit for $f_{\ast,10}$, we see a clear peak for $f_{\ast,10}$ at ${\rm log_{10}}(f_{\ast,10})\sim-2.0$. This is driven by the overlap of the resultant posteriors for the two distinct models outlined in the previous section (i.e. `cold' reionisation and the matter density fluctuations).

For $t_{\ast}$, the disfavoured limits are driven by their degeneracy with the soft-band X-ray luminosity\footnote{Again, this degeneracy only exists in the absence of other constraints on $t_{\ast}$, such as from observed UV galaxy LFs \citep[e.g.][]{Park:2019}.}. In this model the number of X-ray photons produced are inversely proportional to the SFR time-scale. For extremely short SFR time-scales, this can result in a large number of X-ray photons, subsequently heating the IGM and decreasing the amplitude of the resultant 21-cm PS. As a result, this places a disfavoured limit on short star-formation time-scales.

\subsection{Comparison against existing observations}

Now that we have explored the disfavoured astrophysical parameters above, we now shift our focus towards how these models compare against existing observational constraints on the reionisation epoch.

\subsubsection{Reionisation history}

In the middle right panel of Figure~\ref{fig:Corner}, we explore the reionisation histories of these disfavoured astrophysical models. In the shaded region, we present the full range of reionisation histories inconsistent with the latest MWA upper limits on the 21-cm signal. Overlaid on this, we highlight all existing constraints on the IGM neutral fraction. These include limits from the dark pixel statistics of high-$z$ quasars (QSOs; \citealt{McGreer:2015p3668}), the Ly$\alpha$ fraction \citep{Mesinger:2015p1584}, the clustering of Ly$\alpha$ emitters (LAEs; \citealt{Sobacchi:2015}), the Ly$\alpha$ equivalent width distribution of Lyman-break galaxies (LBGs; \citealt{Mason:2018,Hoag:2019,Mason:2019}), the neutral IGM damping wing imprint from high-$z$ QSOs \citep{Greig:2017,Davies:2018,Greig:2019} and the midpoint of reionisation ($z_{\rm Re}$) from Planck \citep{Planck:2018}. Note that this shaded region represents {\it only} those models disfavoured by (i.e. exceed) the recent MWA upper limits. The vast majority of astrophysical models consistent with the existing observational constraints produce 21-cm PS which are below the existing upper limits from MWA.

Importantly, since the range of disfavoured reionisation histories is completely consistent with existing observational constraints, this implies that the latest MWA upper limits on the 21-cm signal are already sufficient to disfavour models which would otherwise be consistent with existing observations. That is, the MWA is already providing unique constraining power on the astrophysics of reionisation. However, owing to the still large amplitude limits, this constraining power is extremely weak. This differs from the picture presented in the equivalent analysis of the LOFAR upper limits \citep{Greig:2020}, where the disfavoured models were already inconsistent with constraints on the IGM neutral fraction. The primary reason for this difference is the fact that the recent MWA limits are at multiple redshift (significantly broadening the range of disfavoured reionisation histories). For example, the models currently disfavoured by LOFAR exceed the upper limits at $z\approx9.1$, whereas the models currently disfavoured by the MWA exceed the upper limits anywhere between $z=6.5-8.7$.

\subsubsection{UV luminosity functions}

In the top right panel of Figure~\ref{fig:Corner}, we show 500 randomly drawn UV LFs (thin black curves) from the posterior of models disfavoured by the MWA upper limits from \citet{Trott:2020} against observations of unlensed UV LFs at $z=6-8$ (orange circles; \citealt{Bouwens:2015,Bouwens:2017}) and at $z=10$ (pink squares; \citep{Oesch:2018}). This shows that the vast majority of UV LFs disfavoured by the current MWA limits are inconsistent with existing UV galaxy limits by several orders of magnitude. In fact, most of these would be strongly ruled out by existing UV LF observations owing to the relatively small observational uncertainties. Nevertheless, there are still some model UV LFs disfavoured by the current MWA limits consistent with these existing observations, again highlighting that the \citet{Trott:2020} limits are providing additional (albeit very weak) constraining power.

It is at this point we emphasise the importance of using \cmfst{} in our Bayesian forward modelling approach. The in-built parameterisation of the ionising sources is able to directly output UV LFs which are capable of being compared to existing observational data. Additionally, this further highlights the synergy between observations of the cosmic 21-cm signal and galaxy UV LFs. Existing galaxy observations already place relatively tight constraints on the UV LF bright end, whereas observing the cosmic 21-cm signal can provide limits on both the very faint end of the underlying UV galaxy LFs and to much higher redshifts. However, we note that interpreting results from the 21-cm signal into galaxy UV LFs is entirely model dependent.

\subsubsection{Electron scattering optical depth, $\tau_{e}$}

\begin{figure} 
	\begin{center}
	  \includegraphics[trim = 0.2cm 0.8cm 0cm 0cm, scale = 0.57]{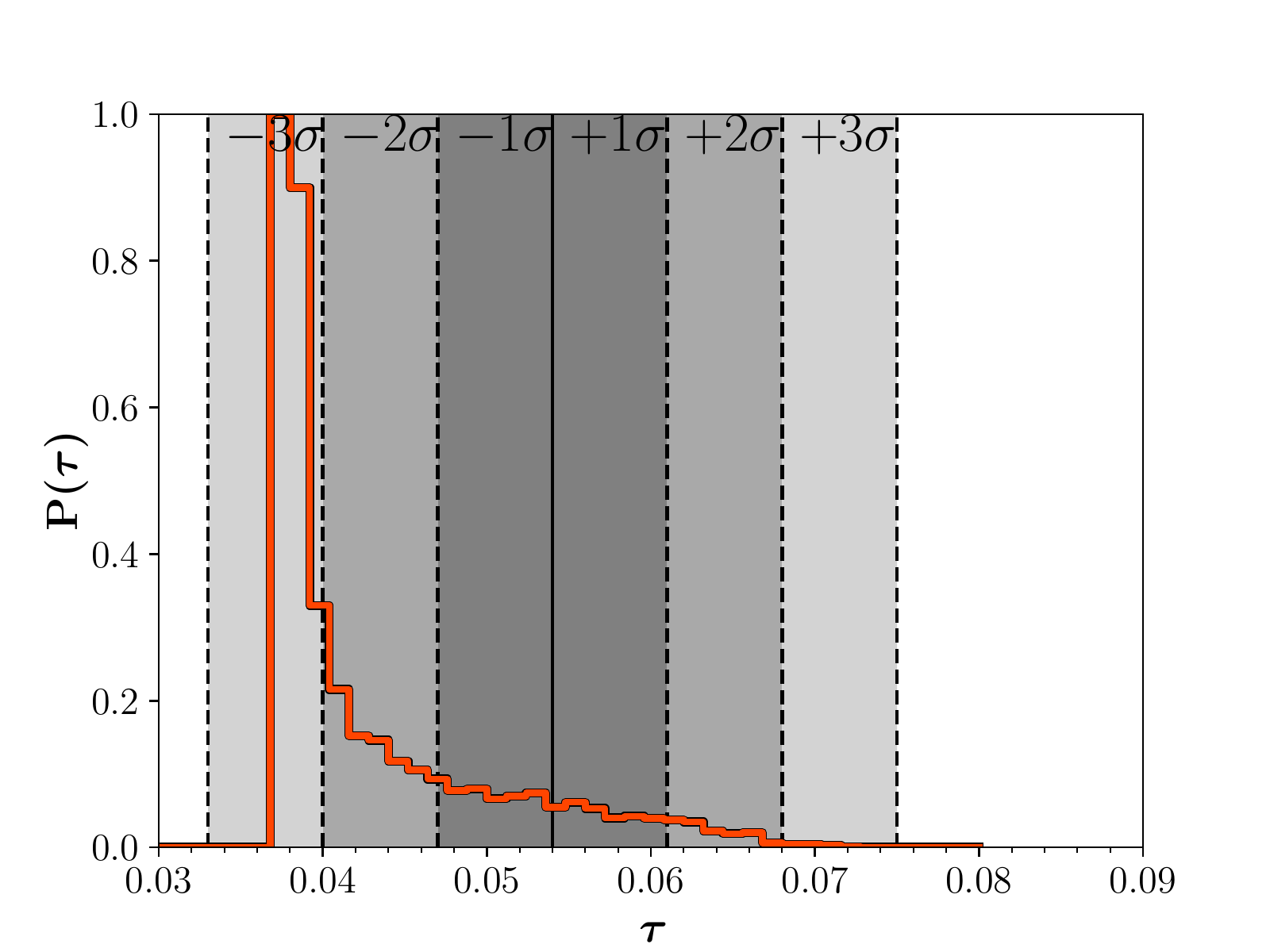}
	\end{center}
\caption[]{Histogram (red curve) of $\tau_{\rm e}$ from all astrophysical models found to be in excess of at least one of the latest MWA upper limits on the 21-cm PS \citep{Trott:2020}. Shaded bands correspond to the statistical uncertainty on $\tau_{\rm e}$ as measured by Planck ($\tau_{\rm e} = 0.054 \pm 0.007$; \citealt{Planck:2018}).}
\label{fig:Tau}
\end{figure}

In Figure~\ref{fig:Tau}, we compare the electron scattering optical depth, $\tau_{\rm e}$, from the models disfavoured by the latest MWA upper limits \citep{Trott:2020} to the latest constraints measured by Planck ($\tau_{\rm e} = 0.054 \pm 0.007$; \citealt{Planck:2018}). The solid vertical line denotes the mean value from Planck, with dashed vertical lines (and shaded regions) denoting $\pm1\sigma$, $\pm2\sigma$, $\pm3\sigma$ from the mean value. The red curve represents a histogram of the $\tau_{\rm e}$ calculated from all models in excess of at least one of the current MWA upper limits.

The vast majority of the models disfavoured by the latest MWA upper limits are inconsistent with existing observations at $\gtrsim2\sigma$. These models prefer very low $\tau_{\rm e}$ values, consistent with the preference for reionisation occurring as late as possible to exceed the MWA upper limits at $z=6.5$ (as discussed in Section~\ref{sec:astro}). It is these models that produce the upper envelope of the shaded region for the disfavoured reionisation histories shown in Figure~\ref{fig:Corner}. Interestingly, the $\tau_{\rm e}$ histogram exhibits an extremely long tail toward larger $\tau_{\rm e}$, crossing the observational constraints from Planck. Again, this implies there are reionisation models consistent with existing observational constraints that are already being disfavoured by the existing MWA upper limits (albeit extremely weakly).

\subsubsection{Mean UV photo-ionisation rate, $\bar{\Gamma}_{\rm UVB}$}

\begin{figure} 
	\begin{center}
	  \includegraphics[trim = 0.1cm 0.8cm 0cm 0.5cm, scale = 0.57]{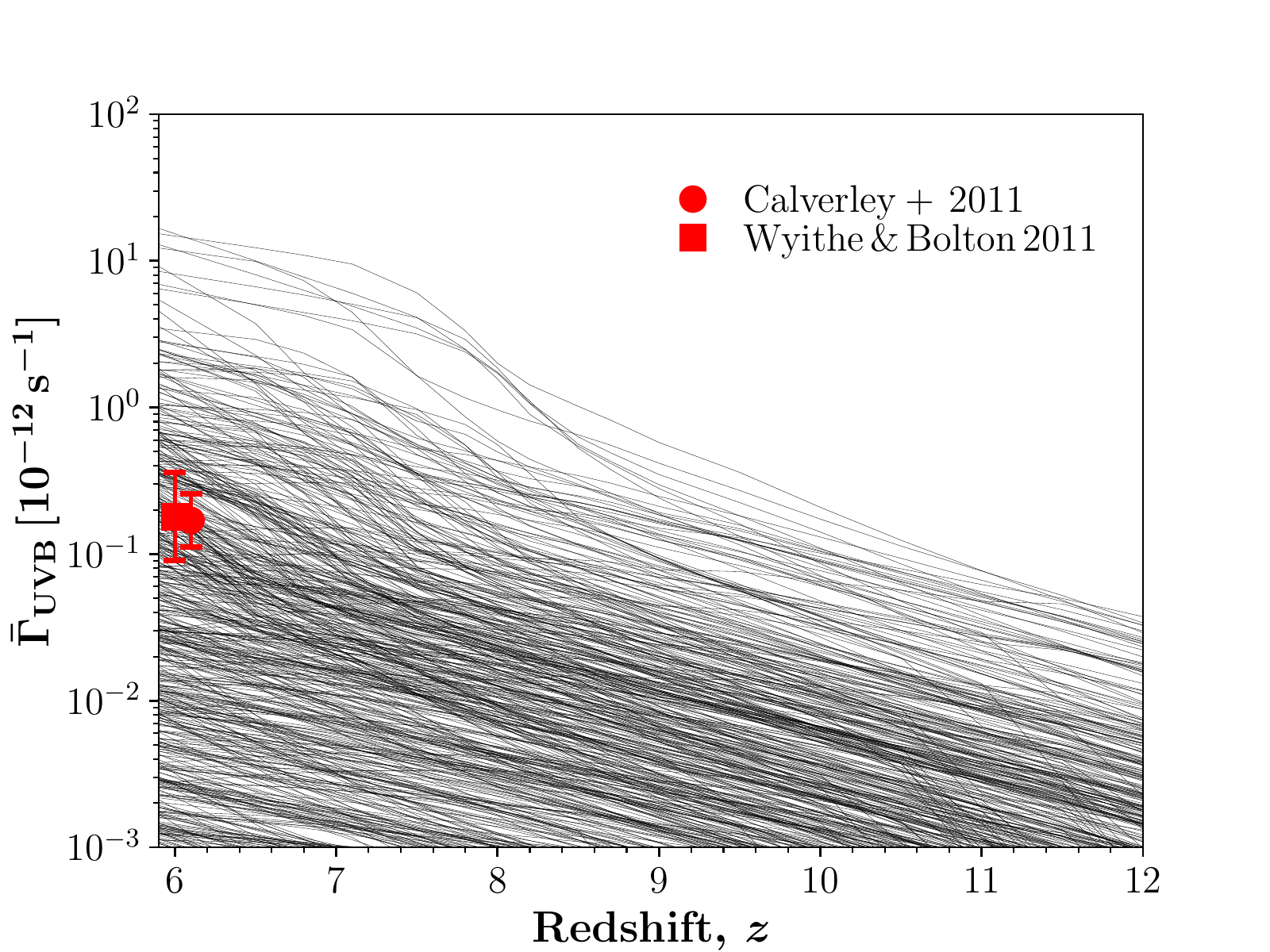}
	\end{center}
\caption[]{A comparison of the mean UV background radiation ($\bar{\Gamma}_{\rm UVB}$) from a random sample of 500 astrophysical models (black curves) found to be in excess of at least one of the recent MWA upper limits on the 21-cm PS \citep{Trott:2020} against observed constraints from the proximity zones of high-$z$ quasars \citep{Calverley:2011,Wyithe:2011}.}
\label{fig:Gamma}
\end{figure}

Finally, in Figure~\ref{fig:Gamma} we compare against the observational constraints on the mean UV background photoionisation rate (red data points) extracted from the proximity zones of $z>6$ QSOs \citep[e.g.][]{Calverley:2011,Wyithe:2011}. Again, we present 500 models (thin black curves) randomly drawn from the posterior of models disfavoured by the latest MWA upper limits \citep{Trott:2020}. Once again, the vast majority of these disfavoured models are already ruled out by existing observational constraints, as highlighted by the large scatter (of several orders of magnitude) in the mean UV background photoionisation rate obtained from the \cmfst{} simulations. Further, the disfavoured models tend to lie on average below the existing observational constraints, implying a reduced output of ionising photons. This is consistent with what we have established from the previous sections. That being, in order to be able to exceed the current MWA upper limits, reionisation is preferred to occur at lower redshifts (or not at all), resulting in a smaller mean photo-ionisation background due to the lower numbers of ionising photons being produced. 

\subsubsection{X-ray emissivity, $\epsilon_{{\rm X},0.5-2~{\rm keV}}$} \label{sec:Xrays}

\begin{figure} 
	\begin{center}
	  \includegraphics[trim = 0.1cm 0.8cm 0cm 0.5cm, scale = 0.57]{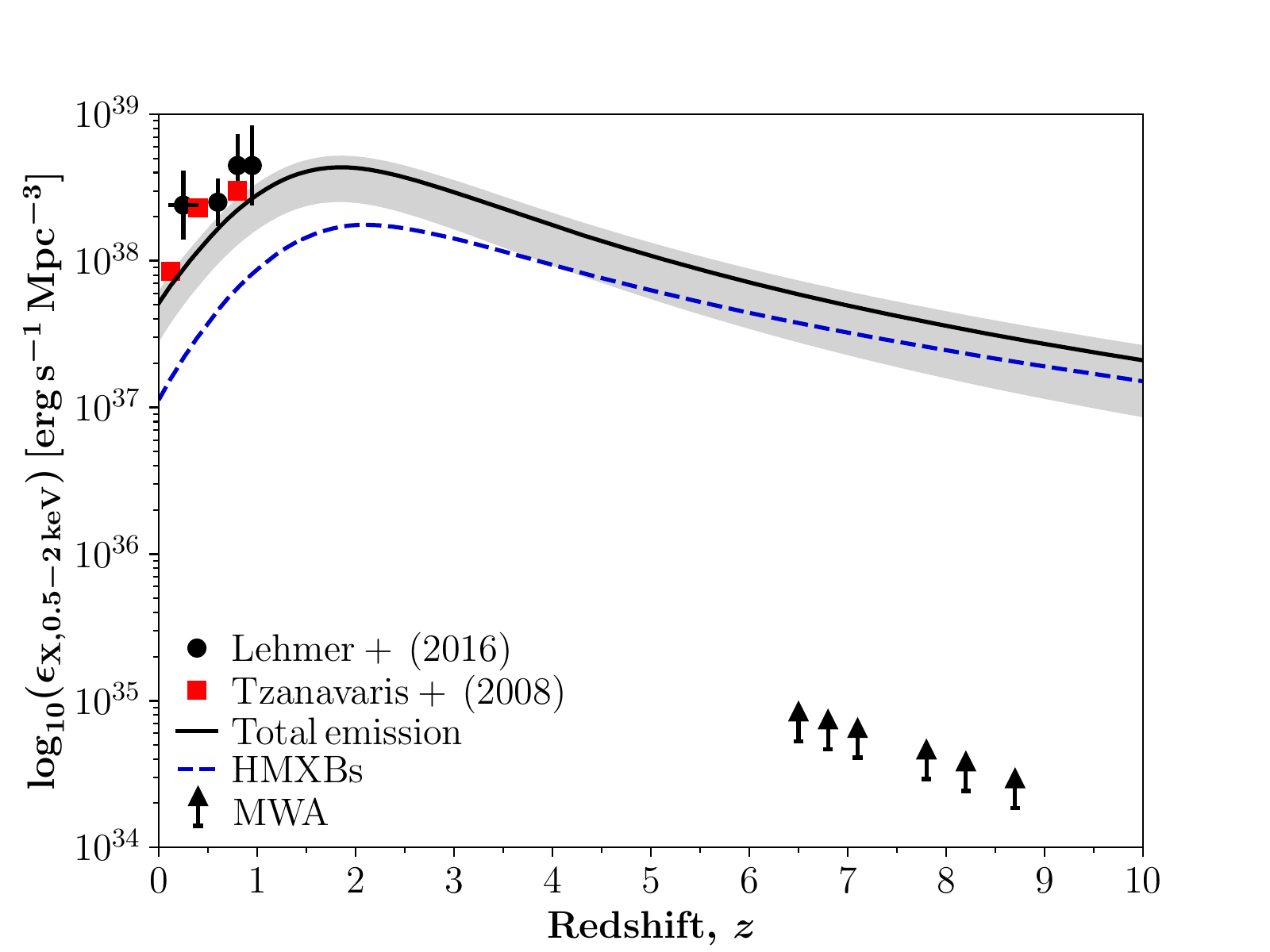}
	\end{center}
\caption[]{A comparison of the soft band (0.5 -- 2 keV) X-ray emissivity, $\epsilon_{{\rm X},0.5-2~{\rm keV}}$, for high redshift galaxies. The lower limits at $z=6.5-8.7$ correspond to the $1\sigma$ limits on the X-ray emissivity inferred from the recent MWA upper limits on the 21-cm PS \citep{Trott:2020}. Observational data is obtained with {\it Chandra}; \citet{Tzanavaris:2008} (red squares) and \citet{Lehmer:2016p7810} (black circles). All curves are obtained from \citep{Lehmer:2016p7810}. The solid curves corresponds to theoretical best-estimates for the evolution of the total X-ray emissivity, with contributions from low-mass X-ray binaries (LMXBs), high-mass X-ray binaries (HMXBs) and a hot interstellar medium. The blue dashed curve corresponds to the HMXB contribution, obtained by scaling Model 269 from \citep{Fragos:2013p6529} to estimates of the stellar mass and SFR density.}
\label{fig:Xray}
\end{figure}

Previously, we have shown that the models disfavoured by the existing MWA upper limits correspond to low X-ray luminosities (i.e. ${\rm log_{10}}\left(L_{{\rm X}<2\,{\rm keV}}/{\rm SFR}\right) \lesssim 37.8$ at 95 per cent confidence). This implies that the neutral IGM is cold, regardless of whether the fluctuations are driven by the patchy EoR or the underlying matter density field (blue and red curves in Figure~\ref{fig:21cmPS}, respectively). As a result, we can place limits on the soft-band X-ray emissivity ($\epsilon_{{\rm X},0.5-2~{\rm keV}}$) of high redshift galaxies (or indeed any source of energy injection into the neutral IGM). At $1\sigma$ we obtain lower limits on the soft band X-ray emissivity from the latest MWA upper limits of ${\rm log_{10}}(\epsilon_{{\rm X},0.5-2~{\rm keV}}) = 34.72, 34.67, 34.61, 34.46, 34.38, 34.27$ at $z = 6.5, 6.8, 7.1, 7.8, 8.2, 8.7$. Note that as these limits are drawn from models that are disfavoured by the current MWA data, thus these limits are conditional on the resultant IGM neutral fraction.

In Figure~\ref{fig:Xray} we compare the resultant lower limits on the soft band (0.5 -- 2 keV) X-ray emissivity implied by the latest MWA upper limits against existing observational constraints. The observational constraints were obtained from large samples of low redshift ($z\leq1$) galaxies obtained with {\it Chandra}; \citet{Tzanavaris:2008} (red squares) and \citet{Lehmer:2016p7810} (black circles). All curves are obtained from \citet{Lehmer:2016p7810}. The solid curve corresponds to the theoretical best-estimates for the evolution of the total X-ray emissivity, with contributions from low-mass X-ray binaries (LMXBs), high-mass X-ray binaries (HMXBs) and a hot interstellar medium. The corresponding $1\sigma$ shaded region accounts for uncertainties in the measurements of the SFR densities and stellar mass densities \citep[e.g.][]{Madau:2014} as well as uncertainties in the XRB SED due to absorption. The blue dashed curve corresponds to the HMXB component, which is expected to dominate at high-redshifts.

Although our limits on the soft band X-ray emissivity are still $\sim$3 orders of magnitude below the expected values, these are the first such limits at these high redshifts. Further, these are obtained from the 21-cm signal, indicating the wealth of information available from the 21-cm signal on the physical properties of the first astrophysical sources. As the lower limits on the 21-cm signal continue to improve, these limits on the X-ray emissivity will approach the theoretical estimates, enabling constraints on the evolution of the X-ray sources to be inferred.

\subsection{Disfavoured IGM properties}

\begin{figure*} 
	\begin{center}
	  \includegraphics[trim = 0.5cm 0.8cm 0cm 0.5cm, scale = 0.72]{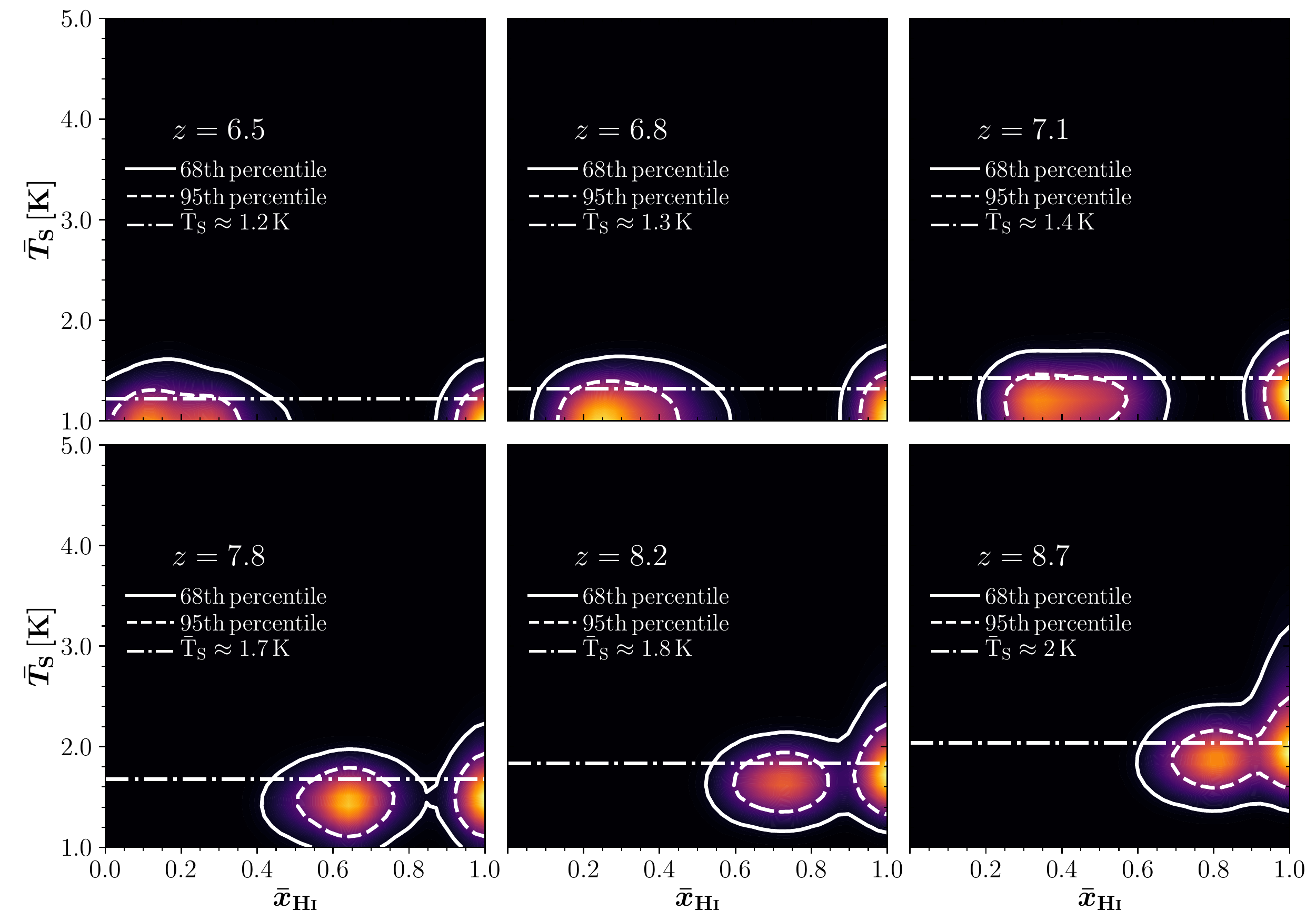}
	\end{center}
\caption[]{2D marginalised posteriors for the IGM neutral fraction, $\bar{x}_{\hi{}}$, and the IGM spin temperature, $\bar{T}_{\rm S}$ for the six different MWA redshift bins presented by \citet{Trott:2020}. Note, these are obtained from our likelihood which combines all six redshift bins simultaneously. Dashed (solid) contours correspond to the 68th (95th) percentile limits.}
\label{fig:TS}
\end{figure*}

After exploring the astrophysical models disfavoured by the latest MWA upper limits \citep{Trott:2020} on the 21-cm signal against existing observations, we now investigate the inferred limits on the globally averaged IGM spin temperature. Note that the IGM spin temperature is not a free parameter of the model, rather it is self-consistently calculated within each simulation voxel within \cmfst{} (see e.g. \citealt{Mesinger:2011p1123}).

In Figure~\ref{fig:TS}, we present the two dimensional marginalised posteriors for the IGM spin temperature, $\bar{T}_{\rm S}$ and the IGM neutral fraction, $\bar{x}_{\hi{}}$ for each of the six redshift limits presented in \citet{Trott:2020}. These two dimensional posteriors are constructed after marginalising the output simulation data by the full posterior of astrophysical model parameters. Dashed (solid) contours correspond to the 68th (95th) percentiles. The vertical dot-dashed line corresponds to the value of the adiabatically cooled neutral IGM calculated at mean density, obtained from \citep[\textsc{RECFAST};][]{Seager:1999p4330,Seager:2000p4329}.

We recover two distinct islands for the $\bar{T}_{\rm S}-\bar{x}_{\hi{}}$ constraints, driven by the two models outlined in Section~\ref{sec:21cmPS}. Firstly, we have an island at $\bar{x}_{\hi{}} > 0.9$, which corresponds to fluctuations solely from the matter density field (i.e. no reionisation). Secondly, an island whose neutral fraction decreases for decreasing redshift as reionisation progresses (the `cold' reionisation scenario). Both islands provide disfavoured limits on the IGM spin temperature, which are driven entirely by astrophysical models that produce little to no heating of the IGM (i.e. ${\rm log_{10}}\left(L_{{\rm X}<2\,{\rm keV}}/{\rm SFR}\right) \lesssim 37.8$ at 95 per cent confidence from Section~\ref{sec:astro}). In the absence of any heating source, the neutral gas in the IGM adiabatically cools with the expansion of the Universe (represented by the vertical dot-dashed line). Note that in Figure~\ref{fig:TS}, the disfavoured regions extend below the limit set by \textsc{RECFAST}, however, this is driven by the fact that \cmfst{} computes the IGM spin temperature on-the-fly in a non-uniform IGM. Due to non-linear structure evolution more of the simulation volume is contained within voids, resulting in the volume averaged IGM spin temperature dropping below the limit set for the neutral IGM at {\it mean} density (the horizontal dot-dashed line).

\begin{figure} 
	\begin{center}
	  \includegraphics[trim = 0.3cm 1.2cm 0cm 0.5cm, scale = 0.34]{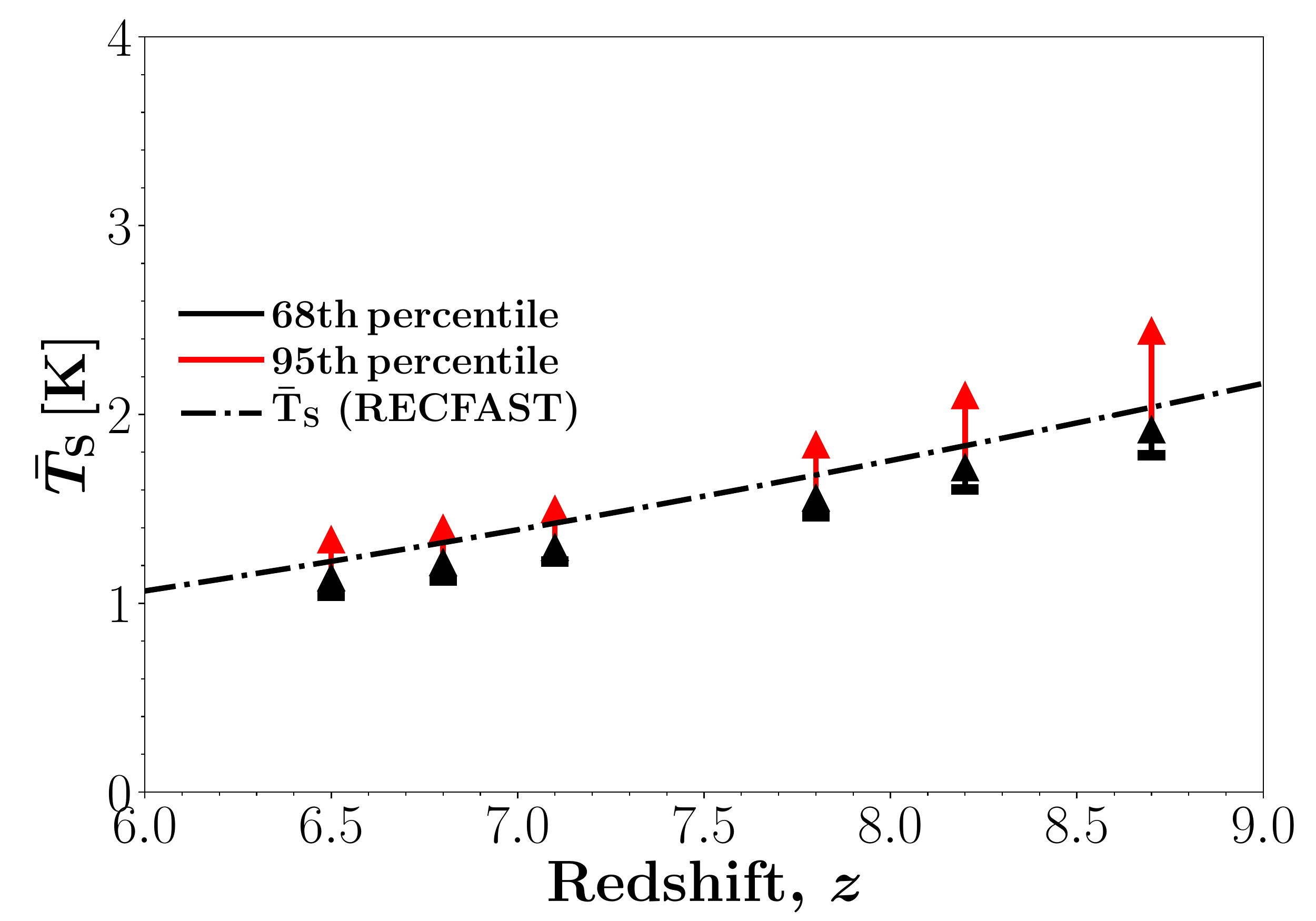}
	\end{center}
\caption[]{Marginalised 1D posteriors on the disfavoured limits on the IGM spin temperature, $\bar{T}_{\rm S}$ for the six different MWA redshift bins presented in \citet{Trott:2020}. These are obtained from our likelihood which combines all six redshift bins simultaneously. Black (red) arrows denote the 68th (95th) percentiles on the disfavoured values of the IGM spin temperature, $\bar{T}_{\rm S}$. The black dot-dashed line corresponds to the value for the neutral IGM at mean density obtained from \textsc{RECFAST}.}
\label{fig:TSz}
\end{figure}

In order to recover disfavoured limits on just the IGM spin temperature, we marginalise our 2D posteriors over the IGM neutral fraction. In Figure~\ref{fig:TSz}, we present the disfavoured limits on the IGM spin temperature for all six redshifts presented in \citet{Trott:2020}. Here, the black (red) arrows denote the 68th (95th) percentiles of the disfavoured limits on the IGM spin temperature from the latest MWA upper limits and the black dot-dashed curve corresponds to the value for a neutral IGM at mean density obtained from \textsc{RECFAST}.

For the six redshift bins with upper limits on the 21-cm signal provided in \citet{Trott:2020}, we recover disfavoured limits on the IGM spin temperature {$\bar{T}_{\rm S}\lesssim$~1.12 (1.32), 1.20 (1.38), 1.28 (1.48), 1.54 (1.82), 1.70 (2.09), 1.90 (2.43)~K at 68th (95th) per cent confidence. As the disfavoured limits extend beyond the adiabatically cooled value for a neutral IGM (dot-dashed curve), this implies that the IGM must have undergone a small amount of heating by X-rays. This is broadly consistent with the interpretation of the recent upper limits from LOFAR at $z\approx9.1$, which also prefer a small amount of X-ray heating (disfavoured limits of $\bar{T}_{\rm S} \lesssim 2.6$ for \citealt{Greig:2020}, $\bar{T}_{\rm S} \lesssim 2.9$ for \citealt{Ghara:2020} and $\bar{T}_{\rm S} \lesssim 10.1$ for \citealt{Mondal:2020} at 95 per cent confidence). Note however that the limits at higher redshifts (i.e. $z > 7.5$) are completely driven by the 21-cm PS upper limits at lower redshifts (i.e. $z=6.5-7.1$), where the 21-cm upper limits are lower in amplitude. This explains the broadening disfavoured regions out to higher redshift, where the variation in the redshift evolution of the IGM spin temperature increases with increasing distance from the better constraining lower redshift upper limits.

\section{Conclusion} \label{sec:Conclusion}

The MWA recently published deep, multi-redshift upper limits at  $z=6.5-8.7$ on the 21-cm PS in the reionisation epoch \citep{Trott:2020}. These were obtained from 298~hours of carefully excised data from four observing seasons, to produce the best upper limits on the 21-cm signal to date at $z<7.5$. At present, these upper limits are still too large to begin to rule out regions of astrophysical parameter space. Instead, following a similar approach to that of the recent analyses of the LOFAR upper limits at $z\approx9.1$ \citep{Ghara:2020,Greig:2020,Mondal:2020}, we explore regions of astrophysical parameter space that are inconsistent with the observational data. We then extend this further to explore how these disfavoured astrophysical models compare against existing observational constraints on the reionisation epoch. We perform this analysis by directly forward modelling the 3D cosmic 21-cm signal using \cmmc{}, an MCMC sampler of 3D reionisation simulations.

We find two classes of astrophysical models disfavoured by the recent MWA upper limits. These are (i) `cold' reionisation models, whereby reionisation proceeds in a cold IGM owing to the lack of a heating source (e.g. no X-ray heating) or (ii) pure matter density fluctuations (i.e. no reionisation). This is the first work to disfavour a signal driven solely by matter density fluctuations.

With respect to the astrophysical parameters, we find that the latest MWA upper limits primarily restrict the soft-band X-ray luminosity of the first galaxies. At 95 per cent confidence, we recover disfavoured limits of ${\rm log_{10}}\left(L_{{\rm X}<2\,{\rm keV}}/{\rm SFR}\right) \lesssim 37.8$. These limits sit below our current expectations from observations of analogue low-redshift star-forming galaxies \citep{Mineo:2012p6282}, stacked {\it Chandra} observations \citep{Lehmer:2016p7810} and predictions at high-redshift from population synthesis models \citep{Fragos:2013p6529}.

In terms of galaxy UV properties, the strongest disfavoured limits are for low values of the parameters controlling the normalisation of the halo mass dependent power-laws for the fraction of galactic gas in stars, $f_{\ast, 10}$ and the ionising escape fraction, $f_{{\rm esc},10}$ and also the star-formation time-scale, $t_{\ast}$. In order to exceed the existing MWA upper limits, the model 21-cm PS needs to be near its peak, which qualitatively occurs roughly around the mid-point of reionisation. Since the lowest amplitude upper limits are achieved at (i.e. $z=6.5$), reionisation must be at its mid-point near $z=6.5$, which can only be achieved by minimising the number of ionising photons available for reionisation (i.e. low $f_{\ast, 10}$ and $f_{{\rm esc},10}$). For $t_{\ast}$, the limits arise due to the degeneracy with the X-ray luminosity. The number density of X-ray photons is inversely proportional to the star-formation time-scale, thus in-order to minimise the amount of X-ray heating to produce large amplitude 21-cm signals, we strongly disfavour short star-formation time-scales.

Next, we extended our exploration to compare the astrophysical models disfavoured by the latest MWA upper limits to existing observation constraints on the reionisation epoch. We compared against: (i) a census of constraints and limits on the IGM neutral fraction, (ii) observed UV galaxy LFs, (iii) the electron scattering optical depth, (iv) the mean UV background photoionisation rate and (v) the soft-band X-ray emissivity. For all, we found that the vast majority of astrophysical models disfavoured by the existing MWA upper limits were already inconsistent with existing constraints. However, we found a small sample of models which were consistent with existing constraints. This implies that the MWA is already bringing unique constraining information to the astrophysics of reionisation, albeit extremely weakly. Using these latest MWA upper limits on the 21-cm signal, we were able to infer the first ever limits on the X-ray properties of galaxies at high redshifts. For the soft-band X-ray emissivity, conditional on the IGM neutral fraction, we recover $1\sigma$ lower limits of ${\rm log_{10}}(\epsilon_{{\rm X},0.5-2~{\rm keV}}) = 34.7, 34.7, 34.6, 34.5, 34.4, 34.3$ at $z = 6.5, 6.8, 7.1, 7.8, 8.2, 8.7$.

Finally, we explored the conditional limits on the IGM spin temperature, $T_{\rm S}$ with the IGM neutral fraction. At 95 per cent confidence, we recover disfavoured limits of $\bar{T}_{\rm S}\lesssim$~1.3, 1.4, 1.5, 1.8, 2.1, 2.4~K at $z=6.5, 6.8, 7.1, 7.8, 8.2, 8.7$, respectively. These limits are found to be above the limits placed by a neutral IGM at mean density undergoing purely adiabatic expansion, implying that the IGM must have undergone some level of X-ray heating. This picture is consistent with that found by \citet{Ghara:2020} and \citet{Greig:2020}, for the recent LOFAR upper limits.

This exploration showcases the value of tools such as \cmmc{}, which forward model the cosmic 21-cm signal in a fully Bayesian framework. In doing so, we are able to infer information about the astrophysics of reionisation from observations of the 21-cm signal. However, note that the astrophysical interpretations within this work are specific to our astrophysical parameterisation and underlying model assumptions. In the near future, as the upper limits on the 21-cm signal continue to improve, we will soon be able to begin to rule out currently viable regions of astrophysical parameter space. Specifically the MWA in the near term should focus on reducing the amplitude of the 21-cm power at the largest scales, (i.e. $k=0.14~h$~Mpc) and lowest redshifts ($z\sim6.5$), to maximise the total disfavoured parameter volume.

\section*{Acknowledgements}

We thank Andrei Mesinger for comments on an early version of this draft. We also thank Bret Lehmer for providing the X-ray emissivity data. Parts of this research were supported by the Australian Research Council Centre of Excellence for All Sky Astrophysics in 3 Dimensions (ASTRO 3D), through project number CE170100013. CMT is supported by an ARC Future Fellowship under grant FT180100321. The International Centre for Radio Astronomy Research (ICRAR) is a Joint Venture of Curtin University and The University of Western Australia, funded by the Western Australian State government. Parts of this work were performed on the OzSTAR national facility at Swinburne University of Technology. OzSTAR is funded by Swinburne University of Technology.

\section*{Data Availability}

The data underlying this article will be shared on reasonable request to the corresponding author.

\bibliography{Papers}

\end{document}